\documentclass{mem}
\usepackage{natbib}\usepackage{txfonts}\usepackage{balance}
\usepackage{graphicx}
\usepackage[a4paper]{hyperref}
\idline{00}{001}
\begin{document}

\def\teff{$T\rm_{eff }$}
\def\kms{$\mathrm {km s}^{-1}$}
\newcommand{\fnl}{f_{\rm NL}}
\newcommand{\bfg}  {\begin{figure}}
\newcommand{\efg}  {\end{figure}}
\newcommand{\incgr} {\includegraphics}
\newcommand{\mbf} {\mathbf}
\newcommand{\be} {\begin{equation}}
\newcommand{\ee} {\end{equation}}
\newcommand{\de} {\,\mathrm{d}}
\newcommand{\bea} {\begin{eqnarray}}
\newcommand{\eea} {\end{eqnarray}}

\title{
Non-Gaussianity in WMAP 5-year CMB map seen through Needlets
}

\subtitle{}

\author{
D. \,Pietrobon\inst{1,2} 
}

\offprints{D.~Pietrobon}

\institute{
University of Roma ``Tor Vergata'',
Via della Ricerca Scientifica 1, 00133 Roma, Italy
\and
Institute of Cosmology and Gravitation,
Dennis Sciama Building
Burnaby Road
Portsmouth, PO1 3FX
United Kingdom\\
\email{davide.pietrobon@roma2.infn.it;davide.pietrobon@port.ac.uk}
}

\authorrunning{Pietrobon}

\titlerunning{NG seen through Needlets}

\abstract{The cosmic microwave background radiation is supposed to be
Gaussian and this hypothesis is in good agreement with the recent very
accurate measurements. Nonetheless a tiny amount of non-Gaussianity is
predicted by the standard inflation scenario, while more
exotic models suggest a higher degree of
non-Gaussianity. Tightly constraining the level of Gaussianity in the
CMB data represents then a fundamental handle to understand the
physics and the origin of our universe. By means of needlets, a novel
rendition of wavelets, characterised by excellent properties of
localisations both in harmonic and pixel domain, we are able to detect
anomalous spots in the southern hemisphere responsible for roughly
the 50\% of power asymmetry we measure in the CMB power spectrum, and to perform
a detailed analysis of the needlets bispectrum. We then constrain the
primordial non-Gaussianity parameter, $\fnl=21\pm40$ at 68\% c.f., and
spot a high asymmetry in the bispectrum, in particular in the
isosceles configurations.

}
\maketitle{}

\section{Introduction}
During the last decade the quality of the cosmological observations
has increased impressively, providing theorists with very accurate
datasets. Cosmic Microwave Background
radiation (CMB) measurements
\citep{Hinshaw2008WMAP5}, SuperNovae IA \citep{Riess:2009pu} and Baryonic Acoustic Oscillations
\citep{Percival:2009xn} fit pretty well within the scenario of 
the so-called Cosmological Concordance Model ($\Lambda$CDM),
which describes fairly well the evolution of the Universe by means of
an handful of parameters \citep{Komatsu2008wmap5}. We live in a flat
universe, whose critical density is provided by baryons
$\Omega_{\rm b}\simeq4\%$, dark matter $\Omega_{\rm c}\simeq25\%$, responsible for the
structures formation, and the troublesome vacuum energy,
$\Omega_\Lambda\simeq70\%$. The other parameters set the normalisation,
$A_{\rm s}$, and
the power, $n_{\rm s}$, of the primordial cosmological fluctuations and the contribution of the stars formation
at more recent epoch, $\tau$. Despite its simplicity, the $\Lambda$CDM
model lacks of a solid theoretical basis. The value and the
origin of the vacuum energy is far from being understood \citep{Copeland:2006}, and the
mechanism which seeds the cosmological perturbations, namely
inflation \citep{Guth1981}, remains more a scenario than a full tested
theory. Our Universe is supposed to be homogeneous and isotropic with
nearly Gaussian fluctuations. Indeed this is in good
agreement with the most recent data, even though the mechanism
generating the perturbations itself should introduce a small non-Gaussian contribution
\citep{Bartolo2004NGreview}. Moreover, exotic early universe scenarios, such as
brane-inspired models \citep{SteinhardtTurok2002}, multi-fields
\citep{LythWands2002} or ekpyrotic models \citep{Mizuno2008} predict a
higher level of non-Gaussianity. Accurately determining the statistics
of the cosmological perturbations represents a fundamental handle on
the early universe physics necessary for understanding the nature and the
evolution of our Universe.

In the following we tackle the issue of non-Gaussianity in the CMB
data, analysing the WMAP 5-year temperature data by means of
needlets. Needlets are a novel rendition of wavelets introduced first
in functional analysis \citep{NarcowichPetrushevWard2006} and then
extended to statistical \citep{Baldi2006}
 and CMB data analysis \citep{Pietrobon2006ISW}.
 Needlets are a basis (more
properly a \emph{tight frame}) defined directly on the sphere, which shows an exponential localisation property both in pixel and harmonic
space. Moreover they are very weakly correlated: this makes them
particularly suitable for CMB data analysis where partial sky
coverage, noise and beam effects have to be minimised in order to
extract tiny signals, especially when non-Gaussian. Needlets are
a quadratic combination of spherical harmonics, weighted by a window
function, $b(\ell/B^j)$
\be
\psi_{jk}(\hat\gamma) = \sqrt{\lambda_{jk}}\sum_{\ell}b\Big(\frac{\ell}{B^{j}}\Big)\sum_{m=-\ell}^{\ell}\overline{Y}_{\ell m}(\hat\gamma)Y_{\ell m}(\xi_{jk}),
\label{eq:needlet}
\ee
where $\hat\gamma$ represents a direction in the sky, $\{\xi_{jk}\}$ is
a set of cubature points (for practical purpose identified with a
pixelization of the sphere \citep{Gorski2005Healpix}), and $B$ a user
chosen parameter which determines the filter width in harmonic
space. For a detailed discussion see \cite{Marinucci2008} and refs.~ therein.
An example of this filter is given in
Fig.~\ref{fig:b_func}, together with the needlet profile in real
space. The localisation property is clearly visible. 
\bfg[htbd]
\center
\incgr[width=.8\columnwidth]{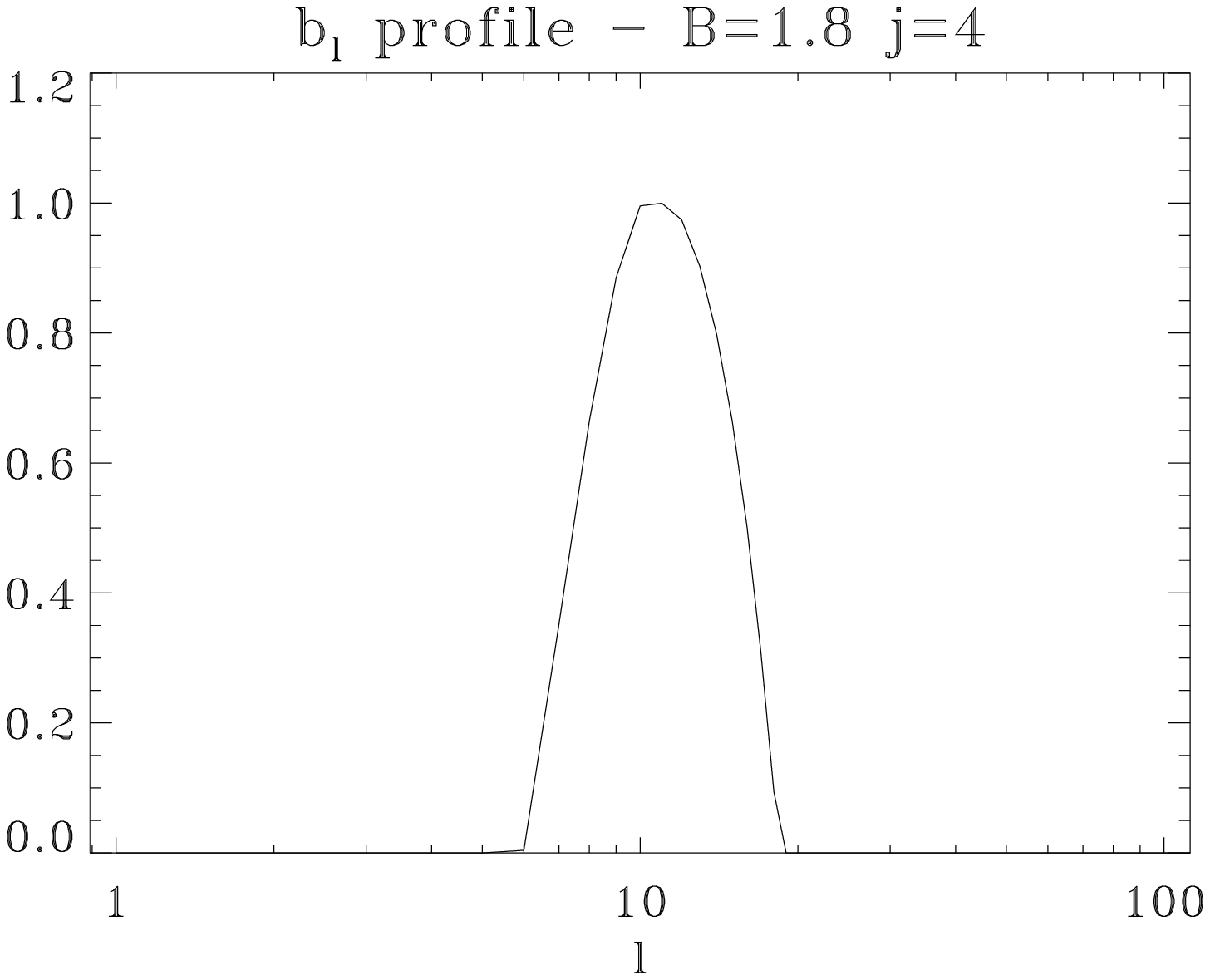}
\incgr[width=.8\columnwidth]{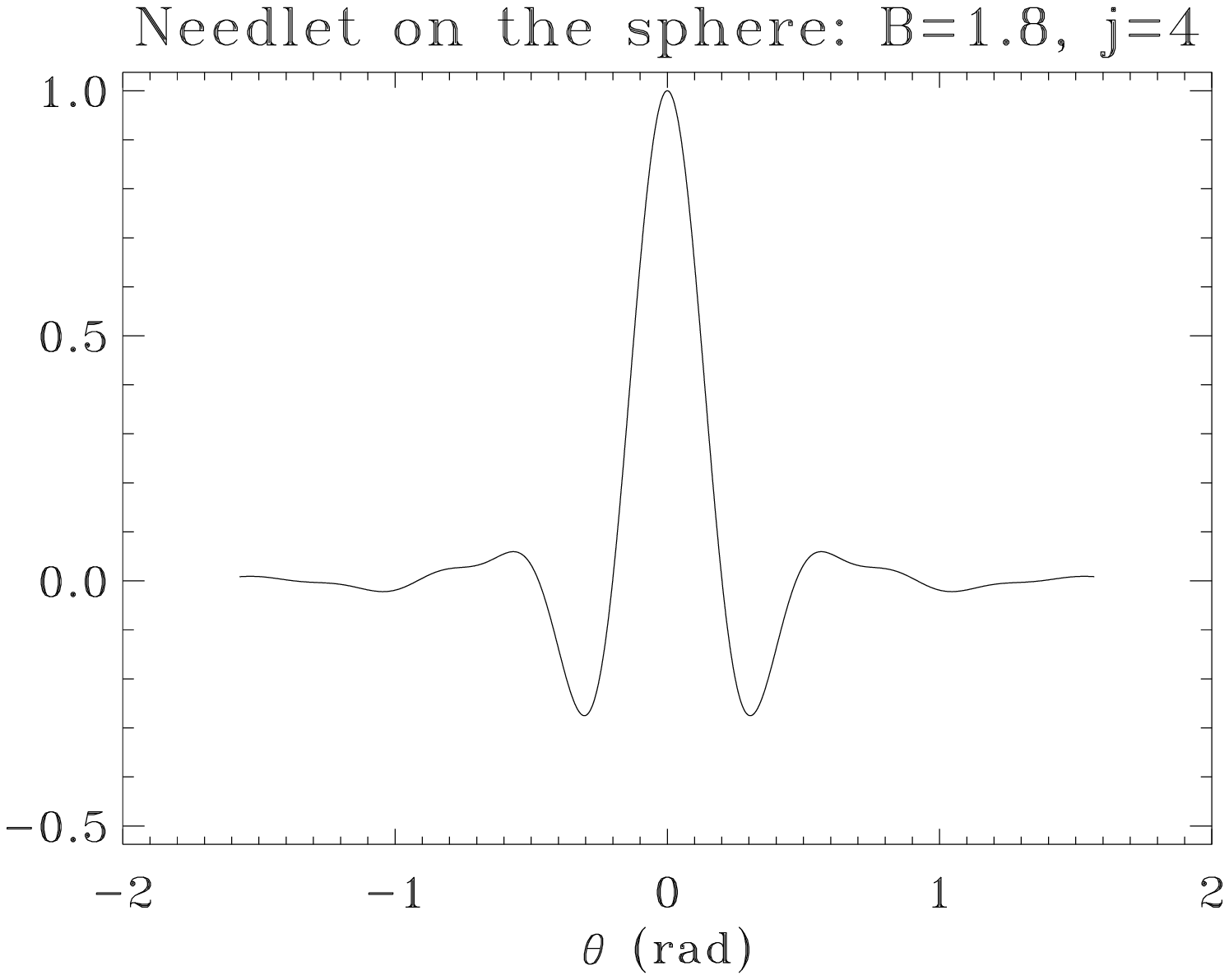}
\caption[Needlet profiles]{\small Profile of the $b(\ell/B^j)$ filter (top)
and needlet (bottom) for the choice $B=1.8$ $j=4$.}
\label{fig:b_func}
\efg
\section{Spot detection and Power spectrum asymmetry}

\begin{figure}[htbd]
\begin{center}
\incgr[width=.25\textwidth, angle=90]{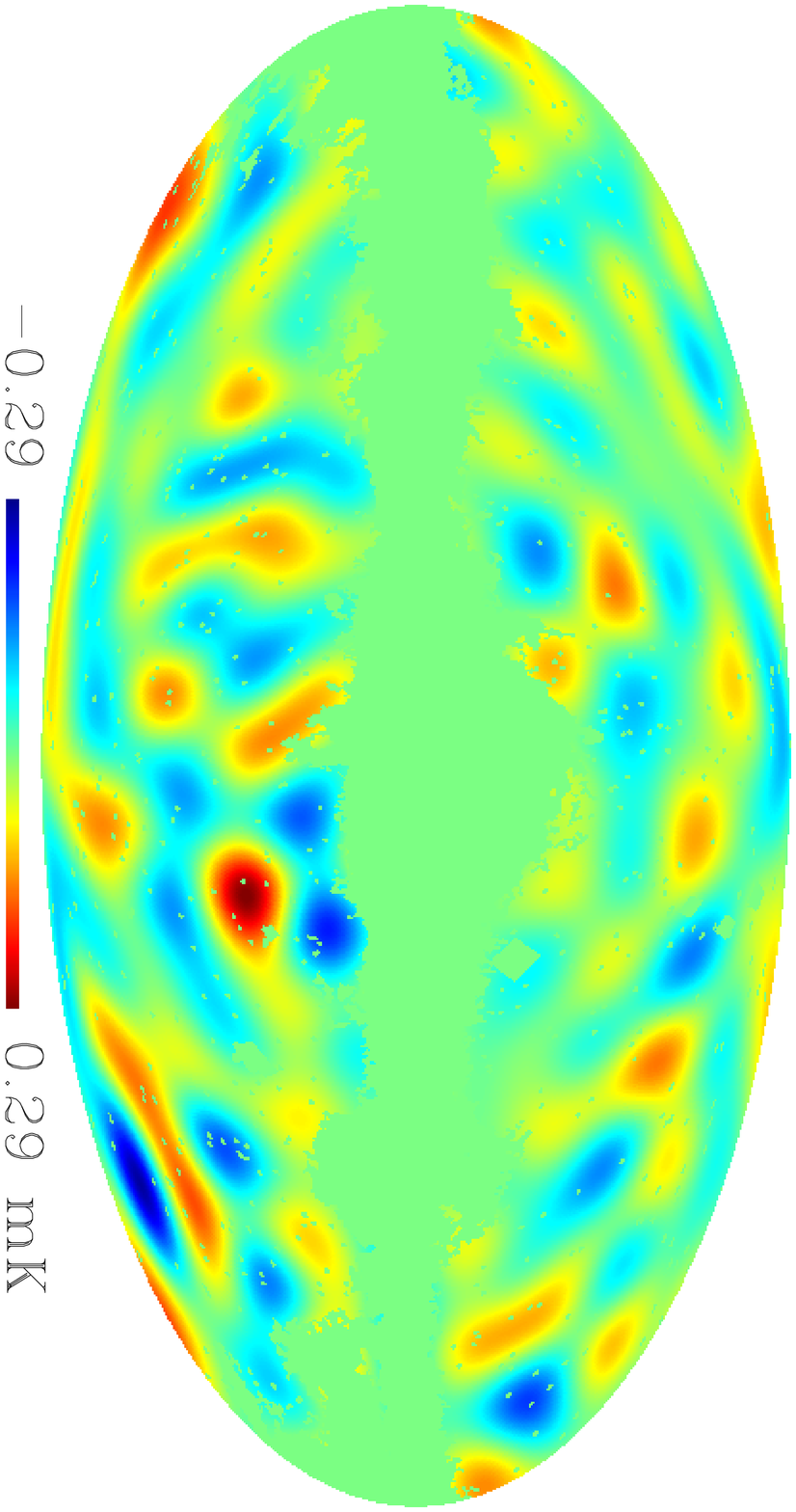}
\caption[Spot detection]{\small Needlet coefficients of the WMAP
5-year CMB temperature map. $B=1.8$ and $j=4$.}
\label{fig:need_coeff}
\end{center}
\end{figure}

Since the first release, the WMAP data have been tested against
Gaussianity and asymmetry (see e.g.
\cite{Eriksen2004,Cruz:2004ce,Land:2006bn,deOliveiraCosta:2003pu}). We applied
needlets to the 5-year data to study the anomaly/asymmetry problem in a coherent
framework. We extracted the needlets
coefficients from the temperature map given by
\be
\beta_{jk} = \sum_\ell b\left(\frac{\ell}{B^{j}}
\right)\sum_m a_{\ell m}Y_{\ell m}(\xi_{jk}). \nonumber
\label{eq:needcoef}
\ee
$\beta_{jk}$ can be easily represented in a mollweide projection: the case of the
Internal Linear
Combination map\footnote{http://lambda.gsfc.nasa.gov/$\sim$/ilc\_map\_get.cfm}
is given in Fig.~\ref{fig:need_coeff}.
Notice that needlets
coefficients are analytically linked to those of the spherical harmonics:
this guarantees an expression for the 2- and 3-point correlation
functions, respectively the power spectrum and bispectrum. By applying needlets, we found three very significant spots in the
southern hemisphere (the brightest features in Fig.~\ref{fig:need_coeff}) which seem to be barely compatible with the
Gaussian hypothesis \citep{Pietrobon2008AISO}. Thanks to needlets localisation, we were able to determine which angular scales they
span. In particular computing the needlets power spectrum
\be
\beta_j\equiv\frac{1}{N_{\rm pix}}\sum_k\beta_{jk}^2=\sum_\ell b^2\Big(\frac{\ell}{B^j}\Big)\frac{2\ell+1}{4\pi}C_\ell,
\label{need_ps}
\ee
where $C_\ell=\langle |a_{\ell m}|^2\rangle$, we found that
the significant power asymmetry between the north and the south
hemisphere is localised at large scales, where the spots
peak. Moreover, masking the spots, we measured a decrease of the
difference in power of a factor 2, underlining the substantial contribution
given by these anomalous features. Fig.~\ref{fig:betaj} summarises our
results. We checked whether this effect influences the
parameter estimation, finding that, with the current precision, the
difference remains within one sigma.
\bfg[htbd]
\center
\incgr[width=.8\columnwidth]{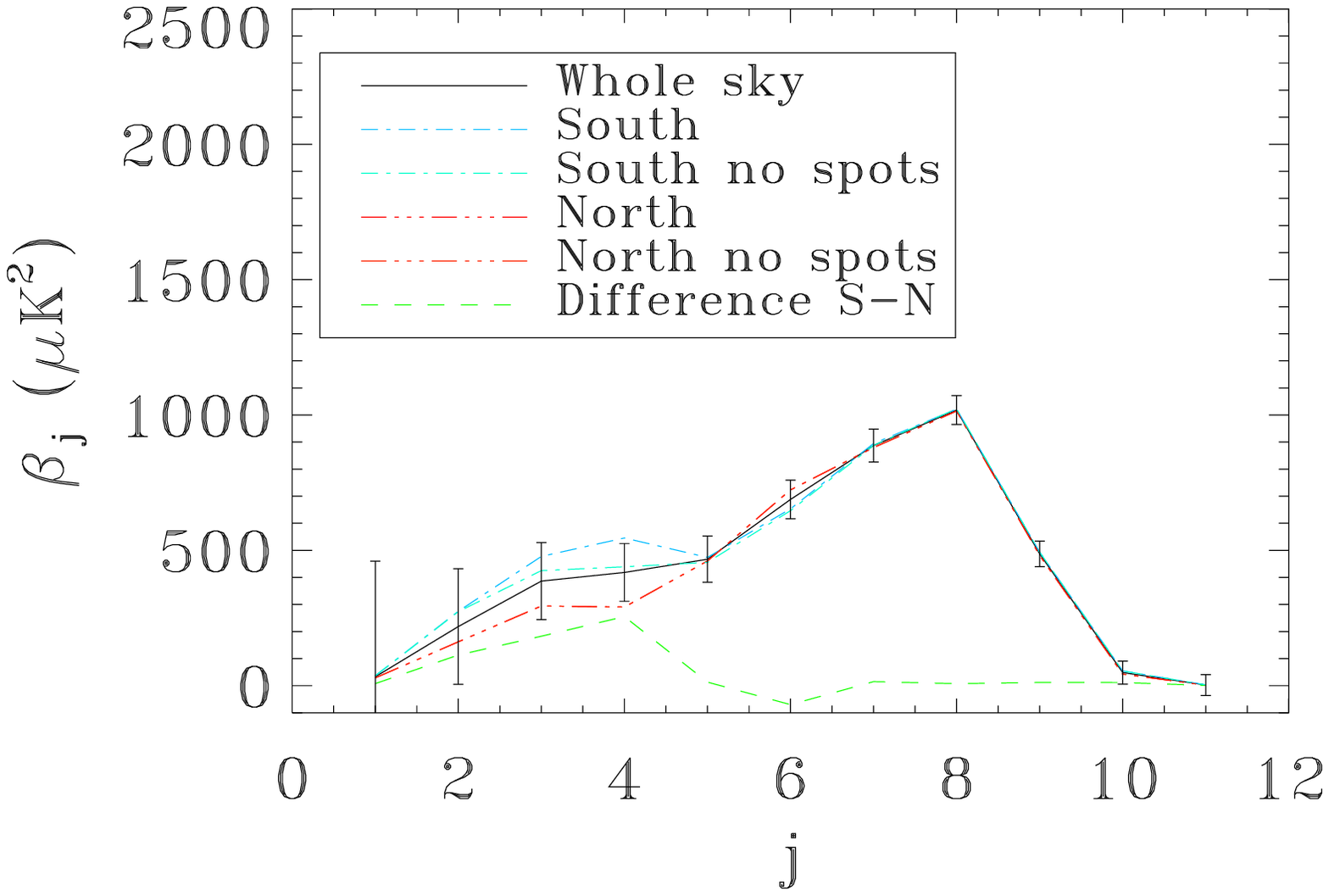}
\caption[Needlet coefficients]{\small $\beta_j$ extracted from WMAP data for different sky cuts. The black solid line shows the signal for the whole CMB sky (kq85 is applied). The blue and red dashed lines show how the power is split between the two hemispheres. When the cold and hot spots detected are masked the excess of power in the southern region is decreased (light blue and orange dot-dashed lines).}
\label{fig:betaj}
\efg

\section{Bispectrum analysis}
The signature of non-Gaussianity appears in the higher moments of a
distribution, which are no longer completely specified by the first
moment (i.e.~the mean value of the distribution) and the second moment
(i.e.~the standard deviation). For a Gaussian distribution, all odd
moments are vanishing, while the even ones can be expressed in term of
just the first two. We then look for a non-vanishing bispectrum of the
distribution of the needlets coefficients. The amplitude of the bispectrum is usually parameterised
by the non-linear parameter $\fnl$ which governs the amplitude of
the non-Gaussian contribution to the primordial gravitational
potential expansion with respect to the linear leading bit:
\be
\Phi(x)=\Phi_{\rm L}(x)+\fnl \left[ \Phi^2_{\rm L}(x) -\langle
\Phi_{\rm L}^2(x)\rangle \right].
\ee
The primordial fluctuations are converted into CMB ones, whose
bispectrum is given by $\langle
a_{\ell_1m_1}a_{\ell_2m_2}a_{\ell_3m_3}\rangle$. In needlets space the
former expression reads:
\be
S_{j_1j_2j_3} =\frac{1}{N_{\rm pix}}
\sum_k\frac{\beta_{j_1k}\beta_{j_2k}\beta_{j_3k}}{\sigma_{j_1}\sigma_{j_2}\sigma_{j_3}},
\label{eq:need_bis}
\ee
where $N_{\rm pix}$ is the number of pixels outside the applied mask
and $\sigma_j$ the variance of the needlets coefficient at the $j$ resolution. See \cite{Pietrobon:2009qg} and Refs.~ therein for a detailed
discussion.
We extracted needlets coefficients from the noise weighted
combination of the WMAP5 channels and measured $\fnl$ by applying the
following estimator \citep{Pietrobon2008NG}:
\be
   \fnl = \frac{\sum_{jj^\prime}S^{\rm obs}_j\mbf{Cov}^{-1}_{jj^\prime}S^{\rm th}_{j^\prime}}{\sum_{jj^\prime}S^{\rm th}_j\mbf{Cov}^{-1}_{jj^\prime}S^{\rm th}_{j^\prime}}.
   \label{eq:fnl_estimator}
\ee
We obtained $\fnl=21\pm40$ at 68\% c.l. We calibrated our result by means
of both Gaussian and non-Gaussian \citep{Liguori2007NGMaps} simulations of the underlying
$\Lambda$CDM best fit model. Our constraints are consistent with those
obtained with different techniques: having several tools, affected by
different systematics, is crucial for analysing the upcoming
cosmological experiments datasets.

The $\fnl$ parameter encodes all the information contained in the
bispectrum irrespective of the specific triangle configuration. The
bispectrum signal can be actually split according to the geometry in
\emph{equilateral}, \emph{isosceles}, \emph{scalene} and \emph{open}
configurations. Different kinds of non-Gaussianity may result in
different shapes, so it is worth looking at them separately. This
is what we addressed in \cite{Pietrobon:2009qg}. The main result of the
paper is summarised in Tab.~\ref{tab:asym_chi2}, where the percentage
of random Gaussian simulations with a $\chi^2$ higher than the data is quoted. 
\begin{table}[htbd]
\begin{center}
\label{tab:asym_chi2}
\begin{tabular}{cccc}
\hline
{\bf conf.}  & {\bf FULL SKY} & {\bf NORTH} & {\bf SOUTH} \\
\hline
\hline
all (115) & $29\%$ & $\mbf{96\%}$ & $\mbf{2\%}$ \\
\hline
equi (9)  & $20\%$ & $11\%$ & $45\%$ \\
\hline
iso (56) & $5\%$ & $\mbf{96\%}$ & $\mbf{0.5\%}$ \\
\hline
scal (50) & $60\%$& $\mbf{90\%}$ & $\mbf{7\%}$ \\
\hline
open (50) & $3\%$& $\mbf{85\%}$ & $\mbf{2\%}$ \\
\hline
\end{tabular}
\end{center}
\caption[\% of the simulations with a $\chi^2$ larger than WMAP5]{\small Percentage of the simulations with a $\chi^2$ larger than WMAP 5-year data
 for the different triangular configurations of the needlets bispectrum.
}
\end{table}
Again, the southern hemisphere results very anomalous, especially in the
isosceles configurations, those characterised by a local type of non-Gaussianity.

\section{Conclusions}
We have discussed the importance of characterising the statistical
distribution of the cosmological perturbations to understand the early
universe physics. We study in detail several properties of the CMB sky
where a non-Gaussian signal may arise by means of needlets. We found anomalous spots in the
southern hemisphere which account for the 50\% of the power spectrum
asymmetries. Moreover we measured the needlets bispectrum constraining
the primordial non-Gaussianity parameter and studying its properties
according to the geometry of the triangle configurations. The
isosceles ones turn out to be the most anomalous: whether this is a
signature of a peculiar early universe model is an interesting issue
in light of the new upcoming experiments.
 
\bibliographystyle{aa}
\bibliography{Bsait}

\end{document}